# Chirality-induced Antisymmetry in Magnetic Domain-Wall Speed

Dae-Yun Kim[1], Min-Ho Park[1], Yong-Keun Park[1,2], Joo-Sung Kim[1], Yoon-Seok Nam[1], Duck-Ho Kim[1,3], Soong-Geun Je[1,4], Byoung-Chul Min[2], and Sug-Bong Choe[1,†]

[1]Department of Physics and Institute of Applied Physics, Seoul National University, Seoul, 08826, Republic of Korea.

[2]Center for Spintronics, Korea Institute of Science and Technology, Seoul, 02792, Republic of Korea.

[3]Institute for Chemical Research, Kyoto University, Kyoto, 611-0011, Japan

[4]CNRS, SPINTECH, F-38042 Grenoble, France

[†]Correspondence to: sugbong@snu.ac.kr

**In chiral magnetic materials, numerous intriguing phenomena such as built-in chiral magnetic domain walls (DWs) [1, 2] and skyrmions [3, 4] are generated by the Dzyaloshinskii–Moriya interaction (DMI) [5, 6]. The DMI also results in asymmetric DW speed under in-plane magnetic field, which provides a useful scheme to measure the DMI strengths [7]. However, recent findings of additional asymmetries such as chiral damping [8-12] have disenabled unambiguous DMI determination and the underlying mechanism of overall asymmetries becomes under debate. By extracting the DMI-induced symmetric contribution, here we experimentally investigated the nature of the additional asymmetry [7]. The results revealed that the additional asymmetry has a truly antisymmetric nature with the typical behavior governed by the DW chirality. In addition, the antisymmetric**

**contribution changes the DW speed more than 100 times, which cannot be solely explained by the chiral damping scenario. By calibrating such antisymmetric contributions, experimental inaccuracies can be largely removed, enabling again the DMI measurement scheme.**

**Introduction**

Magnetic DW motion is of importance, because of its potential application toward the spintronic devices such as memory and logic devices [13-16]. For better performance of these devices, it is crucial to achieve fast DW speed. It has been reported that fast DW speeds can be achieved with built-in chiral DWs [17] generated by the DMI [1, 2], which is an antisymmetric exchange interaction that originated from structural inversion asymmetries [5, 6]. Various experimental techniques have been proposed to quantify the DMI [7, 18-20]. One of the simplest techniques is the symmetry measurement of the DW speed with respect to the in-plane magnetic field [7]. In this technique, the shift of the symmetric axis directly indicates the magnitude of the DMI-induced effective magnetic field. However, many other groups have subsequently reported the existence of the additional asymmetries, which make it difficult to unambiguously determine the symmetric axis. Several possible mechanisms [8-12] have been proposed for the additional asymmetry, but their validities are still under debate. Here, we have experimentally demonstrate the chiral nature of the additional asymmetry. By extracting the well-known symmetric contribution, the additional asymmetry is found to exhibit truly antisymmetric nature, which is governed by the DW chirality.

**Materials and methods**

For this study, 5.0-nm Ta/2.5-nm Pt/0.9-nm Co/2.5-nm X/1.5-nm Pt films were deposited on Si substrates with 100-nm SiO$_2$ layer by DC magnetron sputtering, where X = Al, Pt, W, or Ti. Magnetic wire structures with 20.0-μm width and 500-μm length were then patterned by photolithography technique. All the magnetic wires exhibited strong perpendicular magnetic anisotropy and showed clear domain expansion under the application of an out-of-plane magnetic field $H_z$. The DW images were observed by a magneto-optical Kerr effect microscope equipped with out-of- and in-plane electromagnets. The DW speed $v_{\mathrm{DW}}$ was then measured by analyzing successive DW images captured with a constant time interval under application of $H_z$ as exemplified by Fig. 1(a). This measurement procedure was repeated under application of various current density $J$ and/or in-plane magnetic field $H_x$ bias.

Figure 1(b) plots $v_{\mathrm{DW}}$ with respect to $H_z$ for $+J$ (blue symbols) and $-J$ (red symbols). The two curves in the plot exhibited the same behavior except for the shifts along the abscissa axis, where the direction of the shifts was determined by the polarity of the current bias. When the two curves are shifted by $\pm\Delta H_{\mathrm{eff}}$ along the abscissa axis, they exactly overlap onto a single curve (black open symbols). According to Ref. [21], $\Delta H_{\mathrm{eff}}$ is a direct measure of the effective magnetic field caused by the current bias. The study provides the relation $\Delta H_{\mathrm{eff}} = \varepsilon_{\mathrm{ST}} J$, where $\varepsilon_{\mathrm{ST}}$ is defined as the spin-torque efficiency for the current-to-field conversion [22]. The $\varepsilon_{\mathrm{ST}}$ of various samples was determined from the measurement of $\Delta H_{\mathrm{eff}}$ with respect to $J$ based on the above-mentioned relation. In our experiments, these measurements were obtained for $|J| < 1\times10^{10}$ A/m$^2$, at which the temperature rise due to the Joule heating is negligibly small ($< 0.5$ K) [23].

**Results**

In Fig. 2(a), $\varepsilon_{ST}$ is plotted with respect to $H_x$ for the sample with X = Al. The plot shows the typical spin-orbit-torque-induced behavior composed of three regimes; transient regime (BW-NW regime) in-between two saturation regimes (NW$^\pm$ regimes) [24, 25]. In the NW$^\pm$ regimes, the DW is saturated to the Néel-type configuration with the magnetization inside the DW oriented along the $\pm x$ axes, respectively. On the other hand, in the BW-NW regime, the DW varies between the Bloch- and Néel-type configurations. The $x$ intercept (red vertical line) indicates the magnetic field $H_0$ required to achieve the Bloch-type DW configuration. To achieve such a configuration, $H_0$ should exactly compensate the DMI-induced effective magnetic field $H_{DMI}$, i.e., $H_0 + H_{DMI} = 0$. Therefore $H_{DMI}$ can be quantified from the $x$ intercept of the magnetic field measurements. For this sample, $H_{DMI}$ was estimated as -107.1 mT.

Next, the variation in the DW speed under the application of constant out-of-plane magnetic field $H_z$ with an in-plane magnetic field bias $H_x$ is measured. Figure 2(b) shows $v_{DW}$ with respect to $H_x$. Apart from the symmetric behavior reported in Ref. [7], the present sample exhibited a highly asymmetric behavior. Samples with such large asymmetry need to be analyzed carefully because the field $H_0^*$ (blue vertical line) for minimum $v_{DW}$ occurs far from $H_0$.

Recent studies [7-12] have shown that the DW motion follows the DW creep criticality as given by $v_{DW}(H_x) = v_0 \exp[-\alpha(H_x) H_z^{-1/4}]$, where $v_0$ is the characteristic speed and $\alpha$ is the scaling parameter related to the energy barrier. Here, $\alpha(H_x)$ has the maximum value at the Bloch-type DW configuration and shows symmetric variation with respect to $H_x$ around the maximum. Recalling that the Bloch-type configuration is achieved under application of $H_0$, one can decompose the contribution $S$ symmetric with respect to the axis $H_x = H_0$ as

$$S(\Delta H_x) \equiv \frac{1}{2}[\ln\{v_{\text{DW}}(H_0 + \Delta H_x)\} + \ln\{v_{\text{DW}}(H_0 - \Delta H_x)\}], \quad (1)$$

where $\Delta H_x$ is the deviation of $H_x$ from $H_0$ (i.e. $\Delta H_x \equiv H_x - H_0$). The purple symbols in Fig. 2(c) show $S$ plotted with respect to $\Delta H_x$.

From Eq. (6) in Ref. [7], $\alpha$ can be written as

$$\alpha(H_0 + \Delta H_x) = \begin{cases} \alpha(H_0)\left[1 + \dfrac{2K_D \lambda}{\sigma_0} - \dfrac{\pi \lambda M_S}{\sigma_0}|\Delta H_x|\right]^{1/4} & \text{NW}^{\pm} \text{ Regimes} \\ \alpha(H_0)\left[1 - \dfrac{\pi^2 \lambda M_S^2}{8 K_D \sigma_0}(\Delta H_x)^2\right]^{1/4} & \text{BW} - \text{NW Regime} \end{cases}, \quad (2)$$

where $\lambda$ is the DW width, $M_S$ is the saturation magnetization, $K_D$ is the DW anisotropy, and $\sigma_0$ is the Bloch-type DW energy density. Using the well-known relations $\lambda = \sqrt{A/K_{\text{eff}}}$ and $\sigma_0 = 4\sqrt{AK}$, where $A$ is the exchange stiffness and $K_{\text{eff}}$ is the uniaxial magnetic anisotropy, along with the definitions of the effective anisotropy field $H_K$ ($= 2K_{\text{eff}}/M_S$) and DW saturation field $H_S$ ($= 4K_D/\pi M_S$), Eq. (2) can be written as

$$\alpha(H_0 + \Delta H_x) = \begin{cases} \alpha(H_0)\left[1 - \dfrac{\pi}{4H_K}(2|\Delta H_x| - H_S)\right]^{1/4} & \text{for } |\Delta H_x| > H_S \\ \alpha(H_0)\left[1 - \dfrac{\pi}{4H_K H_S}(\Delta H_x)^2\right]^{1/4} & \text{otherwise} \end{cases}, \quad (3)$$

where the Néel-type configuration in the NW$^{\pm}$ regimes is achieved under the condition $|\Delta H_x| > H_S$.

Because Eq. (3) is expressed using even functions of $\Delta H_x$, the variation in $\alpha$ results in the symmetric contribution $S'$ with respect to $\Delta H_x$ as given by

$$S'(\Delta H_x) \equiv \ln\{v_0(H_0)\} - \alpha(H_0 + \Delta H_x)H_z^{-\frac{1}{4}}. \qquad (4)$$

All the parameters in Eqs. (3) and (4) can be determined by other independent measurements [26] and therefore, we could evaluate $S'(\Delta H_x)$ as shown by the green solid line in Fig. 2(c). It is interesting to see that $S'(\Delta H_x)$ is exactly overlapped onto $S(\Delta H_x)$ even without any fitting parameter in these two curves. This observation implies that the symmetric contribution on the DW speed is mostly attributed to the variation in $\alpha$, which is known to be caused by the variation in the DW energy density.

The additional asymmetry $A$ is then resolved by subtracting the symmetric contribution $S'$ from the DW speed $v_{\mathrm{DW}}$ (i.e. $A \equiv \ln v_{\mathrm{DW}} - S'$). The green symbols in Fig. 2(d) shows $A$ with respect to $H_x$. It is clearly seen from the plot that the additional asymmetry has a truly antisymmetric nature. The negligible deviation (black symbols) from the antisymmetry [27] confirms that no other significant contributions remain.

**Discussion**

The antisymmetry is possibly attributed to either the variation in $v_0$ via the chiral damping mechanism [9, 10] or the additional variation in $\alpha$ via the intrinsic asymmetry of the DW width variation [11]. A clear antisymmetric nature in the logarithmic scale of $v_{\mathrm{DW}}$ indicates the possibility that a large portion of the antisymmetric contribution is rather attributed to the $\alpha$ variation, whereas the $v_0$ variation might cause the asymmetry in linear scale of $v_{\mathrm{DW}}$. Moreover, it is worthwhile to note that the antisymmetric contribution exhibits the typical behavior similar to that of the $\varepsilon_{\mathrm{ST}}$ measurement (Fig. 2(a)) with three regimes composed of transition regime (BW-NW regime) in-between two saturation regimes (NW$^\pm$ regimes). Because the DW chirality governs such typical regimes, one can conclude that the antisymmetric nature of $A$ is mainly attributed to the DW chirality.

Due to the significant antisymmetric contribution, the minimum in $v_{\mathrm{DW}}$ occurs at $H_0^*$ shifted from $H_0$ as shown in Fig. 2 (b), Such apparent $H_0^*$ is often misinterpreted as a compensation field of $H_{\mathrm{DMI}}$. The magnitude of the shift is denoted as $H_{\mathrm{shift}}$ (blue horizontal arrow) in Fig. 2(b). The antisymmetric contribution saturates at $H_{\mathrm{S}}$ and thus, it is natural to expect that $H_{\mathrm{shift}} < H_{\mathrm{S}}$. In addition, $H_{\mathrm{shift}}$ should be proximate to $H_{\mathrm{S}}$ because the antisymmetric contribution varies $v_{\mathrm{DW}}$ more than 100 times, which dominates the symmetric contribution in BW-NW transition regimes. To demonstrate this prediction, $H_{\mathrm{shift}}$ was measured for all the samples with different values of X and then, plotted with respect to $H_{\mathrm{S}}$ as shown by Fig. 3. The black solid line ($H_{\mathrm{shift}} = H_{\mathrm{S}}$) represents the upper bound of $H_{\mathrm{shift}}$. This observation proves that the experimental inaccuracy in the $v_{\mathrm{DW}}(H_x)$-based DMI measurement does not exceed a few tens of mT (as of $H_{\mathrm{S}}$), which might be acceptable for measurement of a large DMI. It is also worthwhile to note that the sign of $H_{\mathrm{shift}}$ is uniquely determined by the DW chirality and therefore, a large portion of the experimental inaccuracy can be removed by calibrating with $H_{\mathrm{S}}$. In Fig. 3, the inaccuracy remains less than about 10 mT. Note that $H_{\mathrm{S}}$ can be estimated as $H_{\mathrm{S}} \cong 2 \ln 2\, \mu_0 M_{\mathrm{S}} t_{\mathrm{f}} / \pi^2 \lambda$, where $t_{\mathrm{f}}$ is the magnetic layer thickness [28].

In summary, the nature of the additional asymmetric contribution on the DW speed variation is examined here. The symmetric contribution of the DW speed is well explained in terms of the DW energy density. The additional asymmetry has a truly antisymmetric nature, which is possibly attributed to the DW chiral chirality. By understanding the nature of antisymmetric contribution, it is possible to remove a large portion of the experimental inaccuracy in the DMI measurement based on the DW-speed asymmetry.

**Acknowledgements**

This work was supported by grants from National Research Foundations of Korea (NRF) funded by the Ministry of Science, ICT and Future Planning of Korea (MSIP) (2015R1A2A1A05001698 and 2015M3D1A1070465). Y.-K.P. and B.-C.M. were supported by the National Research Council of Science & Technology (NST) (Grant No. CAP-16-01-KIST) by the Korea government (MSIP). D.-H.K. was supported from Overseas researcher under Postdoctoral Fellowship of Japan Society for the Promotion of Science (Grant No. P16314).

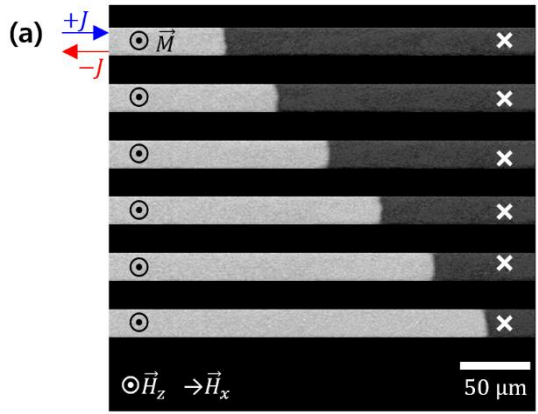

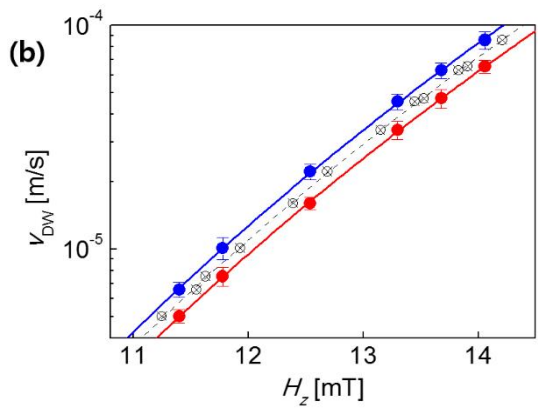

Figure 1. (a) Successive DW images under application of $H_z$ with a constant time interval (1 s). The directions of $H_z$, $H_x$, $J$, and the magnetization $M$ are denoted inside the figure. (b) Plots of $v_{DW}$ with respect to $H_z$ for opposite current bias $+J$ (blue) and $-J$ (red). The black open symbols represent the data for $\pm J$ after it is shifted along the abscissa by $\pm \Delta H_{eff}$, respectively.

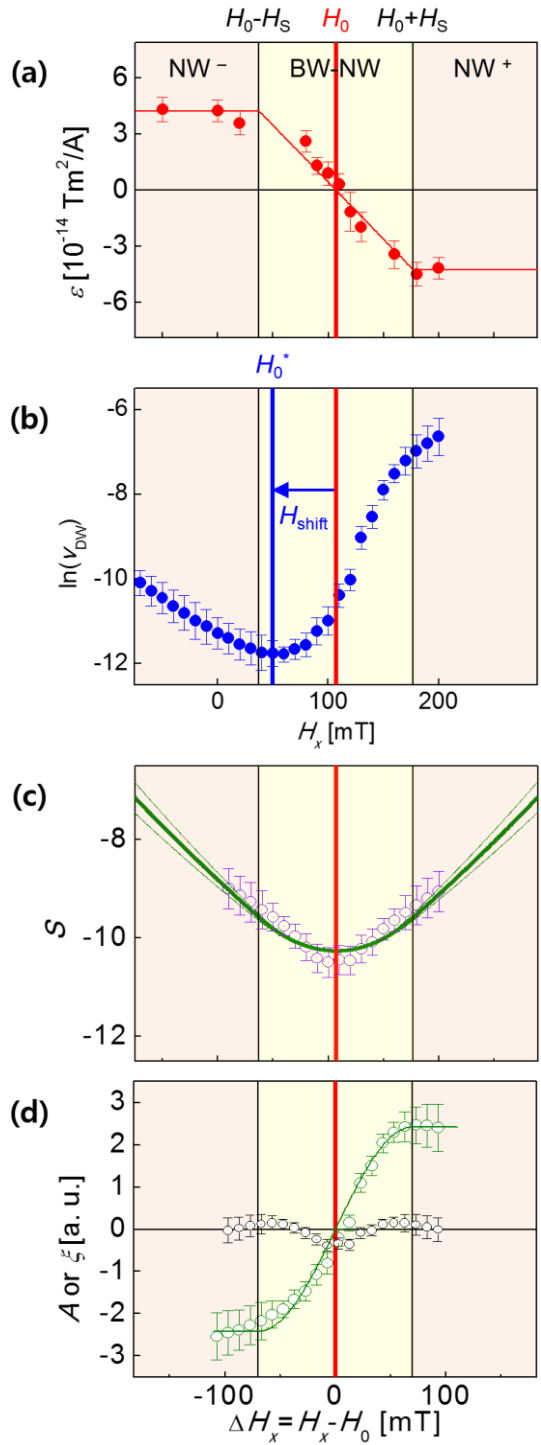

Figure 2. (a) Plot of $\varepsilon_{ST}$ with respect to $H_x$. The red vertical line indicates $H_0$ for the $x$ intercept. The NW$^{\pm}$ and BW-NW regimes are shown using different shaded areas, in which the boundaries are shown by the black vertical lines with the interval $\pm H_S$ from $H_0$. (b) Plot of $v_{DW}$ with respect to $H_x$ under the application of constant $H_z$ (11 mT). The blue vertical line indicates $H_0^*$ for the minimum $v_{DW}$. The blue horizontal arrow shows $H_{shift}$. (c) Plot of $S$

(purple symbols) and $S'$ (green solid line) with respect to $\Delta H_x$. The green dashed lines shows the possible experimental error. (d) Plot of $A$ (green symbols) and $\xi$ (black symbols) with respect to $\Delta H_x$. The green solid line represents antisymmetry.

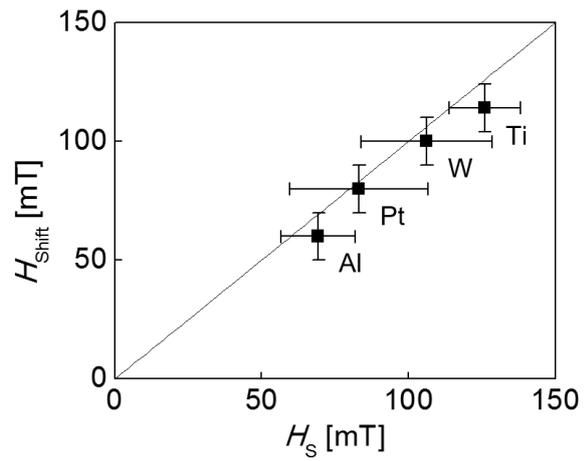

Figure 3. Plot of $H_{shift}$ with respect to $H_S$. The black solid line ($H_{shift} = H_0$) represents the upper bound of $H_{shift}$ for each X.